\documentclass[aps,superscriptaddress,showpacs,twocolumn,nofootinbib]{revtex4}

\usepackage{graphicx}
\usepackage{bm}
\usepackage{amsmath,epsfig}
\usepackage{tabularx}

\newcommand{\abs}[1]{\left\vert#1\right\vert}

\begin{document}



\title{Kullback-Leibler distance as a measure of the information filtered from multivariate data}

\author{Michele Tumminello}
\affiliation{Dipartimento di Fisica e Tecnologie Relative, 
Universit\`a di Palermo, Viale delle Scienze, I-90128 Palermo, Italy}
\affiliation{CNR-INFM, Unit\`a di Palermo, Palermo, Italy}

\author{Fabrizio Lillo}
\affiliation{Dipartimento di Fisica e Tecnologie Relative, 
Universit\`a di Palermo, Viale delle Scienze, I-90128 Palermo, Italy}
\affiliation{Santa Fe Institute, 1399 Hyde Park Road, Santa Fe, NM 87501, U.S.A.}

\author{Rosario N. Mantegna}
\affiliation{Dipartimento di Fisica e Tecnologie Relative, 
Universit\`a di Palermo, Viale delle Scienze, I-90128 Palermo, Italy}

\date{\today}

\begin{abstract}
We show that the Kullback-Leibler distance is a good measure of the statistical uncertainty of correlation matrices estimated by using a finite set of data. For correlation matrices of multivariate Gaussian variables we analytically determine the expected values of the Kullback-Leibler distance of a sample correlation matrix from a reference model and we show that the expected values are known also when the specific model is unknown.  
We propose to make use of the Kullback-Leibler distance to estimate the information extracted from a correlation matrix by correlation filtering procedures. We also show how to use this distance to measure the stability of filtering procedures with respect to statistical uncertainty. We explain the effectiveness of our method by comparing four filtering procedures, two of them being based on spectral analysis and the other two on hierarchical clustering. We compare these techniques as applied both to simulations of factor models and empirical data. We investigate the ability of these filtering procedures in recovering the correlation matrix of models from simulations. We discuss such an ability in terms of both the heterogeneity of model parameters and the length of data series. We also show that the two spectral techniques are typically more informative about the sample correlation matrix than techniques based on hierarchical clustering, whereas the latter are more stable with respect to statistical uncertainty. \end{abstract}

\maketitle

\section{Introduction}
The empirical analysis of interactions between the elements of a complex system is fundamental to understand both the collective structures and the basic rules inducing the emergent behavior of complex systems.
The monitoring of several complex systems nowadays produces large sets of multivariate data. Examples of these sets of data are present in physical \cite{Forrester94, Demasure2003}, biological \cite{maritan,brown, banavar} and economic systems \cite{Laloux1999,Plerou1999,Mantegna1999} and their analysis is an important and challenging task in the investigation of complex systems. Many efforts have been done in the analysis of multivariate data series and most of them focus on the study of pair cross-correlations. 
The analysis of cross-correlation is precious in order to elicit the emergence of collective structures from multivariate data.
Classical spectral methods such as the principal component analysis \cite{Mardia}, recent related techniques based on concepts of random matrix theory \cite{Laloux1999, Plerou1999}, hierarchical clustering \cite{Anderberg}, factor analysis \cite{Mardia} and graph theory \cite{West} are fruitful approaches to the analysis of correlations among elements of complex systems elicited by multivariate data.

Cross-correlations estimated from real data are unavoidably affected by the statistical uncertainty due to the finite size of the sample. In most cases, the length of data is unavoidably limited whereas in other cases the length of data needs to be limited to avoid that sizable non-stationary effects might introduce large errors in the estimation of correlations. Cross-correlations might also be affected by noise due to measurement errors and to the interaction of the system with the environment. In order to at least partially overcome these problems, it is advisable to select statistically reliable information from the correlation matrix. We address the selection of the most statistically reliable part of the correlation matrix with the locution {\it filtering of the correlation matrix}.

Several techniques have been proposed in the literature in order to filter out information from the correlation matrix and therefore it is important to have at hand a method for comparing the performance of such different techniques in a quantitative way.

In this paper, we propose to measure the performance of filtering procedures by using the Kullback-Leibler distance \cite{Kullback} which is a measure of distance between probability distributions and it is widely used in information theory (see for instance \cite{Cover}). Specifically, for multivariate Gaussian variables, we explicitly compute the analytical form of the Kullback-Leibler distance and we show how it depends on the correlation matrices of the considered sets of data or of filtered versions of them. Under the same assumptions we analytically obtain the expected values of the Kullback-Leibler distance between the correlation matrix of a multivariate model and a sample correlation matrix obtained with the Pearson estimator from a finite set of data. One of our key results is that these expected values are model independent. This result shows that the Kullback-Leibler distance is very good in quantifying the amount of information present in a sample correlation matrix with respect to an hypothetical reference model also in the cases when the specific nature of the model is unknown. We are also able to compute the expected value of the Kullback-Leibler distance between two distinct samples of the correlation matrix obtained from the same random source. This last quantity is very useful in quantifying the stability associated with any sample estimation and specifically with the stability of the correlation matrices obtained from filtering procedures.   

We show the effectiveness of the use of the Kullback-Leibler distance in comparing data and models and in assessing the stability of the estimation of the sample correlation matrix by investigating four different filtering methods. Two of them are based on spectral analysis, while the other two are generated by hierarchical clustering procedures. A good filtered correlation matrix is supposed to be informative about the sample correlation matrix and, at the same time, to be statistically more robust than the sample matrix itself with respect to statistical uncertainty. In our investigation we consider in a quantitative way both these aspects.

The paper is organized as follows. In section \ref{kullbackTHEOR} we present the analytical results of the expected values of the Kullback-Leibler distance and we show how the Kullback-Leibler distance can be used as an estimator of the goodness of filtering procedures. In section \ref{methods} we describe the four filtering procedures that we quantitatively compare in section \ref{comparison} both by investigating simulations and real data. Finally, in section \ref{conclusions} we draw our conclusions.    

\section{Kullback-Leibler distance}\label{kullbackTHEOR}

The Kullback-Leibler distance (see for instance \cite{Cover}) or \emph{mutual entropy} is a measure of the distance between two probability densities, say $p$ and $q$, which is defined as
\begin{equation}
\label{kullbackEQGEN}
K(p,q)=E_p\left[\log{\left( \frac{p}{q} \right)}\right],
\end{equation}
where $E_p[\, . \, ]$ indicates the expectation value with respect to the probability density $p$. The Kullback-Leibler distance is asymmetric. In Eq.(\ref{kullbackEQGEN}) the expectation value is evaluated according to the distribution $p$. Since the property of symmetry is sometimes important a symmetrization of the Kullback-Leibler distance, called Jefferys-Kullback-Leibler J-divergence has been introduced \cite{Jefferys, Kullback}. 
In other cases the asymmetry could also be an useful feature of a distance measure. This is the case when objects of different nature (or simply with different statistical meaning) are compared. 
The Kullback-Leibler distance is widely used in information theory. The mutual information between two random variables $X$ and $Y$ is defined as $K(p(X,Y),p(X) p(Y))$ (see for instance \cite{Cover}), where $p(X,Y)$ is the joint probability density function of $X$ and $Y$, whereas $p(X)$ and $p(Y)$ are the corresponding marginal probabilities. In this case, the asymmetry is important because the mutual information is measuring the error one commits in considering  two random variables as independent variables. Accordingly, this measure is performed by evaluating the distance between the correct joint probability $p(X,Y)$ and the product $p(X) p(Y)$, averaging the result over $p(X,Y)$.

Here we consider the Kullback-Leibler distance between multivariate Gaussian random variables. We consider variables with zero mean and unit variance without loss of generality because we are interested in the comparison of the correlation matrices of the two set of variables. In this case, the Gaussian multivariate distribution associated with the random vector $X$ is completely defined by the correlation matrix ${\bf \Sigma}$ of the system. In the following we indicate the probability density function with $P({\bf \Sigma},X)$.
Given two different probability density functions $P({\bf \Sigma_1},X)$ and $P({\bf \Sigma_2},X)$, we have
\begin{widetext}
\begin{equation}
\label{kullbackMulti}
K(P({\bf \Sigma_1},X),P({\bf \Sigma_2},X))=E_{P({\bf \Sigma_1},X)}\left[\log{\left( \frac{P({\bf \Sigma_1},X)}{P({\bf \Sigma_2},X)} \right)}\right]
=\int{P({\bf \Sigma_1},X) \log\left[ \frac{P({\bf \Sigma_1},X)}{P({\bf \Sigma_2},X)}\right] dX},
\end{equation}
\end{widetext}
By performing the integral in Eq. (\ref{kullbackMulti}) one obtains:
\begin{eqnarray}
\label{kullbackGaussian}
K(P({\bf \Sigma_1},X),P({\bf \Sigma_2},X))=\frac{1}{2} \left[\log{\left(\frac{\abs{{\bf \Sigma_2}}}{\abs{{\bf \Sigma_1}}}\right)}+\right. \nonumber \\
\left. +\text{tr}\left({{\bf \Sigma_2}^{-1} {\bf \Sigma_1} }\right)-n\right],
\end{eqnarray}
where $n$ is the dimension of the space spanned by the $X$ variable and ${\abs{{\bf \Sigma}}}$ indicates the determinant of ${\bf \Sigma}$.
In Appendix A we show how to derive the last equation from Eq. (\ref{kullbackMulti}).
Eq. (\ref{kullbackGaussian}) shows that the Kullback-Leibler distance is an explicit function of only the correlation matrices ${\bf \Sigma_1}$ and ${\bf \Sigma_2}$ for multivariate normal distributions. Therefore, from now on we indicate $K(P({\bf \Sigma_1},X),P({\bf \Sigma_2},X))$ simply with $K({\bf \Sigma_1},{\bf \Sigma_2})$. It is worth noting that the Kullback-Leibler distance takes naturally into account the statistical nature of correlation matrices. Indeed $K({\bf \Sigma_1},{\bf \Sigma_2})$ is well defined only provided that the matrices ${\bf \Sigma_1}$ and ${\bf \Sigma_2}$ are positive definite. This property is not common to other measures of distance between matrices which are based essentially on the isomorphism between the matrix space and a vector space, e.g. the Frobenius distance (see below). However this property can also be a limitation. The Kullback-Leibler distance cannot be used to quantify the distance between semi-positive correlation matrices that are observed when  the length $T$ of data series is smaller than the number $n$ of elements of the system.
%
%
The Kullback-Leibler distance is also related to the Maximum Likelihood Factor Analysis (MLFA) \cite{Mardia}. In fact, the log-likelihood function to be maximized in order to describe a system of $n$ elements with sample correlation matrix ${\bf C}$ estimated from data series of length $T$, with a certain \emph{k}-factor model with correlation matrix ${\bf \Sigma_k}$ is given by:
\begin{equation}
\label{loglikekull}
L({\bf C},{\bf \Sigma_k})=-T \, K({\bf C},{\bf \Sigma_k}) -\frac{1}{2} T  \left[ \log{ \left( \abs{2 \pi {\bf C}} \right) }-n\right].
\end{equation}
In the MLFA, $L({\bf C},{\bf \Sigma_k})$ is maximized with respect to ${\bf \Sigma_k}$. This maximization is therefore equivalent to minimize the Kullback-Leibler distance $K({\bf C},{\bf \Sigma_k})$ with respect to ${\bf \Sigma_k}$, because the other terms in Eq. (\ref{loglikekull}) are independent of ${\bf \Sigma_k}$. 
It is to notice that in Eq. (\ref{loglikekull}) the empirical correlation matrix ${\bf C}$ is the one estimated from the investigated data and one calibrates the correlation matrix ${\bf \Sigma_k}$ of the model by maximizing $L({\bf C},{\bf \Sigma_k})$. This fact explains why the log-likelihood is depending on $K({\bf C},{\bf \Sigma_k})$ instead of $K({\bf \Sigma_k},{\bf C})$. 

In this paper we want to apply the Kullback-Leibler distance to sample correlation matrices obtained with the Pearson estimator. Since different realizations of the process give rise to different samples, a Kullback-Leibler distance having one or two sample correlation matrices as arguments is a function of one or two random matrices. It is known that sample covariance matrices of finite variance variables belong to the ensemble of Wishart random matrices and many statistical properties of Wishart matrices are known \cite{Mardia}. It is therefore useful to investigate the statistical properties of Kullback-Leibler distance involving sample correlation matrices of multivariate Gaussian random variables. These properties will be useful in the next section as absolute terms of comparison of filtering procedures of the correlation matrix.

Let us consider a multinormally distributed random vector $X$ of dimension $n$ with correlation matrix ${\bf \Sigma}$. Let ${\bf C_1}$ and ${\bf C_2}$ be two sample correlation matrices obtained from two independent realizations of the system both of length $T$.
By making use of the theory of Wishart matrices \cite{Mardia} we obtain that
\begin{eqnarray}
\label{kullexpecSigS1}
E\left[K({\bf \Sigma},{\bf C_1})\right]=\frac{1}{2} \left \{n\log{\left(\frac{2}{T}\right)}+\right. \nonumber \\
\left. +\sum_{p=T-n+1}^{T}{\left[\frac{\Gamma^{\prime}(p/2)}{\Gamma(p/2)}\right]}+\frac{n (n+1)}{T-n-1}\right\},
\end{eqnarray}
\begin{eqnarray}
\label{kullexpecS1Sig}
E\left[K({\bf C_1},{\bf \Sigma})\right]=\frac{1}{2} \left \{n\log{\left(\frac{T}{2}\right)}-\right. \nonumber \\
\left. -\sum_{p=T-n+1}^{T}{\left[\frac{\Gamma^{\prime}(p/2)}{\Gamma(p/2)}\right]}\right\}
\end{eqnarray}
and
 \begin{equation}
\label{kullexpecS1S2}
E\left[K({\bf C_1},{\bf C_2})\right]=\frac{1}{2} \frac{n (n+1)}{T-n-1},
\end{equation}
where 
%
$\Gamma(x)$ is the usual Gamma function and $\Gamma^{\prime}(x)$ is the derivative of $\Gamma(x)$. In Appendix B we show how to derive these expectation values. 
%
%
Finally, it is possible to give the asymptotic expectation value of the standard deviation of $K({\bf C_1},{\bf \Sigma})$ by using the Bartlett statistics \cite{Bartlett54}. Specifically if $T\gg1$, $n\gg1$ and $Q=T/n\gg1$ we infer that the standard deviation of $K({\bf C_1},{\bf \Sigma})$ is $\sigma_K\simeq1/(2Q)$.\\
It is important to observe that all the expectation values given in Eq.s (\ref{kullexpecSigS1}-\ref{kullexpecS1S2}) are independent of ${\bf \Sigma}$, i.e. they are independent of the specific model. 
This fact implies that (i) the Kullback-Leibler distance is a good measure of the statistical uncertainty of correlation matrix which is due to the finite length of data series and (ii) the expected value of the Kullback-Leibler distance is known also when the underlying model hypothesized to describe the system is unknown. This fact has important consequences. Suppose one 
knows that the observed data are well approximated by a multivariate Gaussian distribution and that one measures a sample correlation matrix ${\bf C}$. In order to remove some unavoidably present statistical uncertainty, the experimenter applies a filtering procedure to the data obtaining the filtered correlation matrix ${\bf C^{filt}}$. If the filtering technique is able to recover the model correlation matrix, i.e. ${\bf C^{filt}}={\bf \Sigma}$, the Kullback-Leibler distance $K({\bf C},{\bf C^{filt}})$ must be equal on average to the value given in Eq.~(\ref{kullexpecS1Sig}). This expected value is independent on the (unknown) model correlation matrix ${\bf \Sigma}$.   
Therefore large deviations from this expectation value indicate that the filtered matrix is not consistent with the true matrix of the system. If $K({\bf C},{\bf C^{filt}})$ is significantly smaller (in terms of the error $\sigma_K\simeq1/(2Q)$) than the expectation value of Eq. (\ref{kullexpecS1Sig}), it means that the filtering procedure has at most partially removed the statistical uncertainty, i.e. the filtered matrix is keeping some of the statistical uncertainty due to the finite length $T$. If, on the other hand, $K({\bf C},{\bf C^{filt}})$ is significantly larger than the value of Eq. (\ref{kullexpecS1Sig}), it means that the filtered matrix is either filtering too much information or distorting the signal. The distance between $K({\bf C},{\bf C^{filt}})$ and the expected value of Eq.~(\ref{kullexpecS1Sig}) is a measure of the goodness of the filtering procedure in keeping the maximal amount of information which can be present in sample correlation matrices estimated with a finite number of records.
   
 A second aspect concerns the stability of the filtered correlation matrix obtained from a sample matrix. Let us suppose to apply a certain filtering procedure to the correlation matrices ${\bf C_1}$ and ${\bf C_2}$ of two independent realizations of the system, obtaining two filtered correlation matrices ${\bf C^{filt}_1}$ and ${\bf C^{filt}_2}$. If it turns out that $K({\bf C^{filt}_1},{\bf C^{filt}_2})$ is larger than the expected value of  $K({\bf C_1},{\bf C_2})$ described by Eq.~(\ref{kullexpecS1S2}), one can conclude that the filtering procedure produces correlation matrices less reproducible than the sample correlation matrices and therefore the procedure is not suitable for the purpose of filtering robust information from the empirical correlation matrices ${\bf C_1}$ and ${\bf C_2}$.

In summary we have shown that the Kullback-Leibler distance is very good for comparing correlation matrices because (i) it is an asymmetric distance and therefore it can distinguish between quantities observed in real systems and used to model the empirical observations, e.g. the sample correlation matrix and the filtered correlation matrix respectively; 
(ii) the expectation values of the Kullback-Leibler distance given in Eq.s (\ref{kullexpecSigS1}-\ref{kullexpecS1S2}) are model independent, indicating that this distance is a good estimator of the statistical uncertainty due to the finite size of the empirical sample; (iii) the Kullback-Leibler distance is intimately related to the log-likelihood function used in MLFA and (iv) it is deeply related with concepts of information theory, such as the the mutual information. These properties are not observed in other widespread distances between matrices. For example, we shall show that we do not find these properties in the Frobenius distance, which is a standard measure of the distance between matrices. 

The Frobenius distance between two $n \times n$ matrices ${\bf \Sigma_1}$ and ${\bf \Sigma_2}$, of real elements $s_{ij}^1$ and $s_{ij}^2$ respectively, is defined as
\begin{eqnarray}
\label{frobenius}
F\left({\bf \Sigma_1},{\bf \Sigma_2} \right) & =\sqrt{\sum_{i=1}^n \sum_{j=1}^n (s_{ij}^1-s_{ij}^2)^2}\nonumber \\
& =\sqrt{\text{tr}\left[\left({\bf \Sigma_1}-{\bf \Sigma_2}\right) \left({\bf \Sigma_1}-{\bf \Sigma_2}\right)^T\right]}
\end{eqnarray}
We note that the Frobenius distance is symmetric. Therefore it cannot assign a different role to a model correlation matrix ${\bf \Sigma}$ with respect to some sample ${\bf C}$ of ${\bf \Sigma}$. We also observe that this distance is well defined independently of the statistical nature of matrices ${\bf \Sigma_1}$ and ${\bf \Sigma_2}$, i.e. these matrices can also be non positive definite. 
Finally and more important, we want to show, for a simple system of two variables, that the expectation value of the Frobenius distance between a true correlation matrix and its Pearson estimator is model dependent, i.e. this expectation value depends on the true correlation matrix.\\

Let us consider a bivariate normal distribution $N(\textbf{0},{\bf \Sigma})$, where ${\bf \Sigma}$ is a $2 \times 2$ correlation matrix and $\textbf{0}$ is the null vector of dimension $2$. We indicate the only entry of ${\bf \Sigma}$ different from 1 with $\rho$. The sample correlation matrix ${\bf C}$ is defined as
\begin{equation}\label{samplematrix}
{\bf C}=
\begin{pmatrix}
  1    &  \hat{\rho}  \\
   \hat{\rho}   &  1
\end{pmatrix},
\end{equation}
where $\hat{\rho}$ is the Pearson correlation coefficient estimated from a realization of $N(\textbf{0},{\bf \Sigma})$ of length $T$. It results that
\begin{equation}
\label{frobenius22}
F\left({\bf \Sigma},{\bf C} \right)=\sqrt{2} \abs{\rho-\hat{\rho}}
\end{equation}
The distribution of $\hat{\rho}$ is approximately Gaussian for large values of $T$. The mean value of $\hat{\rho}$ is $\rho$ and the standard deviation is $(1-\rho^2)/\sqrt{T}$ \cite{Fisz}. Accordingly, the expectation value of the Frobenius distance between the two matrices is:
\begin{equation}
\label{expecfrobenius22}
E\left[F\left({\bf \Sigma},{\bf C} \right)\right]=\frac{2}{\sqrt{\pi \, T}} (1-\rho^2)
\end{equation}
This result shows that the Frobenius distance is model dependent and therefore it is not a good estimator of the statistical uncertainty of correlation matrix due to the finite length of data series.

\section{Filtering procedures}\label{methods}

In this section we describe four procedures that can be used to filter correlation matrices. Two procedures are based on spectral techniques, i.e. they are based on the comparison between the spectrum of the sample correlation matrix and the spectrum expected for a random matrix. These procedures are described in some detail in subsection \ref{spectral}.
The other two techniques that we consider here are hierarchical clustering procedures. Specifically, we obtain two different filtered matrices by applying the Single Linkage Cluster Analysis (SLCA) and the Average Linkage Cluster Analysis (ALCA) to the sample correlation matrix of the system. The ALCA and SLCA are standard procedures of hierarchical clustering and we describe how these techniques generate filtered correlation matrices in subsection \ref{HCtheor}.

\subsection{Spectral methods}\label{spectral}

Random matrix theory \cite{Metha90} was originally developed in nuclear physics and then applied to many different fields. 
Let us consider $n$ independent random variables with finite variance and $T$ records each. The sample correlation matrix of the system in the limit $T\to \infty$ is simply the identity matrix. When $T$ is finite the correlation matrix will in general be different from the identity matrix. Random matrix theory allows to prove that in the limit $T,n \to \infty$, with a fixed ratio $Q=T/n \geq 1$, the eigenvalues of the sample correlation matrix ${\bf C}$ cannot be larger than
%
%
\begin{equation}\label{lmax}
\lambda_{max}=\sigma^2 (1+1/Q+2\sqrt{1/Q}),
\end{equation}  
where $\sigma^2=1$ for correlation matrices. 
The idea underlying both the spectral filtering procedures considered here is that of reducing the impact of eigenvalues smaller than $\lambda_{max}$ on the structure of an empirical correlation matrix, in order to remove the effects of those eigenvalues that are consistent with the null hypothesis of uncorrelated random variables. In some practical cases, such as for example in finance, one finds that the largest eigenvalue $\lambda_1$ of the empirical correlation matrix is definitely inconsistent with random matrix theory. In these cases, the null hypothesis is modified so that correlations can be explained in terms of a one factor model. Accordingly, when $\lambda_1>>\lambda_{max}$ we set $\sigma^2=1-\lambda_1/n$ in Eq. (\ref{lmax}) \cite{Laloux1999}.\\ 
The first filtering procedure we consider here has been used by Rosenow \emph{et al.} in Ref. \cite{Rosenow2002}. The technique consists in replacing the eigenvalues smaller than $\lambda_{max}$ in the diagonal matrix ${\bf D}$ of eigenvalues of ${\bf C}$ with $0$'s,  thus obtaining a new diagonal matrix ${\bf D^*_S}$. One can therefore compute the matrix ${\bf Q_S}= {\bf V^T}\,{\bf D^*_S} {\bf V}$ of elements $q_{ij}^S$, where ${\bf V}$ is the matrix of eigenvectors of ${\bf C}$. Finally, the filtered correlation matrix ${\bf C^S}$ of elements $c_{ij}^S$ is obtained by forcing the diagonal elements of ${\bf Q_S}$ to 1, i.e. $c_{ij}^S=\delta_{ij}+q_{ij}^S\,(1-\delta_{ij})$, where $\delta_{ij}$ is the standard Kronecker symbol. 
The second procedure we apply has been considered by Potters \emph{et al.} in Ref.  \cite{Potters2005}. Here, eigenvalues smaller than $\lambda_{max}$ in ${\bf D}$ are replaced with their average value in the diagonal matrix ${\bf D^*_B}$. As in the previous case, one rotates the matrix ${\bf D^*_B}$ getting the matrix ${\bf Q_B}= {\bf V^T}\,{\bf D^*_B} {\bf V}$ of elements $q_{ij}^B$, where again ${\bf V}$ is the matrix of eigenvectors of ${\bf C}$. Finally, the filtered correlation matrix ${\bf C^B}$ is the matrix of elements $c_{ij}^B=q_{ij}^B/\sqrt{q_{ii}^B\,q_{jj}^B}$ .
%
Both the matrices ${\bf C^S}$ and ${\bf C^B}$ satisfy the properties of a correlation matrix, i.e. (i) they are positive definite; (ii) their diagonal elements are equal to 1 and (iii) their off-diagonal elements are in absolute value smaller or equal to 1.
%
%
%
%
%
\subsection{Hierarchical Clustering Procedures}\label{HCtheor}

Another approach used to filter the information associated with the correlation matrix is given by hierarchical clustering analysis \cite{Anderberg}. Let us consider a set of $n$ objects and suppose that a similarity measure, e.g. the correlation coefficient, between pairs of elements is defined. Similarity measures can be written in a $n\times n$ similarity matrix. The hierarchical clustering methods allow to hierarchically organize the elements in clusters. A result of the procedure is a rooted tree or dendrogram giving a quantitative description of the clusters thus obtained. Another result of the procedure is a filtered correlation matrix. Indeed the whole information about the rooted tree can be stored in a $n \times n$ matrix ${\bf C^<}$ \cite{Anderberg}. We have recently shown \cite{TumminelloEPL2007} that, when the entries of ${\bf C^<}$ are non negative numbers, this matrix is the correlation matrix of a suitable factor model, that we have named Hierarchically Nested Factor Model (HNFM). This result ensures that, under the condition of non negative entries of ${\bf C^<}$ (typically satisfied in many empirical applications), this matrix is a true correlation matrix, i.e. it is positive definite.  

A large number of hierarchical clustering procedures can be found in the literature. For a review about the classical techniques see for instance Ref. \cite{Anderberg}. In this paper we focus our attention on the SLCA and the ALCA.\\

The starting point of both the procedures is the empirical correlation matrix ${\bf C}$.
The following procedure performs the ALCA giving as an output a rooted tree and a filtered correlation matrix ${\bf{C}}^<_{\bf ALCA}$ of elements $c^<_{ij}$:
\begin{enumerate} 
\item Set ${\bf{B}} = {\bf{C}}$.
\item Select the maximum correlation $b_{hk}$ in the correlation matrix ${\bf{B}}$. Note that after the first step of construction $h$ and $k$ can be simple elements (i.e. clusters of one element each) or clusters (sets of elements). $\forall \, i \in h$ and $\forall \, j \in k$ one sets the elements $c^<_{ij}$ of the matrix ${\bf{C}}^<_{\bf ALCA}$ as $c^<_{ij}=c^<_{ji}=b_{hk}$.
\item Merge cluster $h$ and cluster $k$ into a single cluster, say $q$. The merging operation identifies a node in the rooted tree connecting clusters $h$ and $k$ at the correlation $b_{hk}$. 
\item Redefine the matrix ${\bf{B}}$:
\begin{eqnarray} \label{negspin}
\left \{  \begin{aligned}
        &   b_{qj}= \frac{n_h\,b_{hj}+n_k \, b_{kj}}
                               {n_h+n_k}  & 
                           ~~~~\text{ if } j\notin h \,{\rm{and}} \, j\notin k\\
        &                 \nonumber \\
        &    b_{ij}=b_{ij} & 
                           ~~~~\text{ otherwise, }\\
\end{aligned} \right.
\end{eqnarray}
where $n_h$ and $n_k$ are the number of elements belonging respectively to the cluster $h$ and to the cluster $k$ before the merging operation. Note that if the dimension of ${\bf{B}}$ is $m \times m$ then the dimension of the redefined ${\bf{B}}$ is $(m-1) \times (m-1)$ because of the merging of clusters $h$ and $k$ into the cluster $q$.
\item If the dimension of ${\bf{B}}$ is larger than 1 then go to step 2, else Stop.   
\end{enumerate}
By replacing point $4$ of the above algorithm with the following item\\

\indent $4.$  Redefine the matrix ${\bf{B}}$:
\begin{equation}\label{negspin2}
\left \{ \begin{aligned}
        &  b_{qj}= Max \left[b_{hj}, b_{kj}\right] & 
                           ~~~~\text{ if } j\notin h \,{\rm{and}} \, j\notin k \nonumber\\
        &  b_{ij}=b_{ij} & 
                           ~~~~\text{ otherwise, }\\
\end{aligned} \right.
\end{equation}
one obtains an algorithm performing the SLCA and the associated filtered correlation matrix  ${\bf{C}}^<_{\bf SLCA}$. In the following, we indicate the matrices 
${\bf{C}}^<_{\bf SLCA}$ and ${\bf{C}}^<_{\bf ALCA}$ with ${\bf{C^{SLCA}}}$ and ${\bf{C^{ALCA}}}$, respectively, in order to simplify the notation.
    
\section{Comparison of filtering procedures}\label{comparison}

We have applied the four filtering procedures described in the previous section to both real and artificial systems. We have considered the real system of daily returns of the 100 most capitalized stocks traded at New York Stock Exchange (NYSE) in the time period from January 2001 to December 2003. In this case, the length of the $n=100$ time series is $T=748$ records. We have also considered the system of daily returns of 92 highly capitalized stocks traded at London Stock Exchange in 2002. The length of the $n=92$ time series is $T=250$ for this system. We have also applied the filtering procedures to two artificial systems of $n=100$ elements each. Both these systems are described by a factor model \cite{Mardia}. A factor model is a mathematical model which describes the correlation among a set of elements that we indicate with $x_i$ $(i=1,...,n)$, in terms of a certain number of common factors $f_k$ $(k=1,...,P)$. The linear dependence of elements from factors is mathematically expressed as
\begin{equation}
\label{facGEN}
x_i(t)=\sum_{k=1}^P\gamma_{ik}f_k(t)+\eta_i \epsilon_i(t),
\end{equation}
where $i\in\{1,...,n\}$, $\eta_i=[1-\sum_{k=1}^P\gamma_{ik}^{2}]^{1/2}$. The $k^{th}$ factor $f_k(t)$ and $\epsilon_i(t)$ are independent identically distributed random variables with zero mean and unit variance. In our simulations, the factors  $f_k(t)$ $(k=1,...,P)$ and the idiosyncratic noises $\epsilon_i(t)$ $(i=1,...,n)$ are Gaussian random variables. 
%
%

In the first artificial system that we consider here, elements are grouped in $P=12$ orthogonal clusters. In terms of factor models, this orthogonal grouping of elements is expressed by the fact that elements belonging to different clusters depend on different (independent) factors, i.e. if $x_i$ belongs to the group $k$ then $x_i(t)=\gamma_{ik}f_k(t)+\eta_i \epsilon_i(t)$. The dimension of groups is heterogeneous to mimic typical conditions observed in some real systems. Specifically the number of elements belonging to each group ranges from a minimum of 3 elements to a maximum of 17. The other artificial system that we have considered is described by a HNFM with $P=23$ factors. This empirically based model has been introduced in Ref.  \cite{TumminelloEPL2007}.  We have chosen these two models because they are conceptually very different one from the other. In fact, in the HNFM elements cannot be straightforwardly divided in groups because they depend on factors in a nested hierarchical way whereas in the other model the groups of elements are clearly distinguished because elements belonging to different groups depend on different and mutually independent factors. Roughly speaking we can say that the block diagonal model describes a ``separable'' system whereas the HNFM represents a ``nested'' system. In a first analysis, both the considered factor models are degenerate models, i.e. the coefficient $\gamma_{ik}$, which expresses the dependence of the element $i$ on the factor $k$ in the model of Eq. (\ref{facGEN}), is only depending on the factor and not on the element.
It is to notice that by applying either the ALCA or the SLCA to the correlation matrix of the two considered models one obtains back the correlation matrix of the models. This fact is due to the degeneracy of the models and it gives a certain advantage to hierarchical clustering procedures with respect to spectral techniques in reconstructing the true correlation matrix of these systems. In fact both the considered spectral techniques cannot reconstruct the true correlation matrix ${\bf \Sigma}$ of the system when applied to ${\bf \Sigma}$ itself.
This is the first reason why we have decided to perform other simulations of the systems by removing the degeneracy from models. The second reason is that the true correlation matrix of the system is in general unknown for real data: we have only one correlation matrix obtained from a single realization of the system with finite time series length $T$. Accordingly, we have decided to perform one single realization, say ${\bf X_{T_d}}$, with length $T_d$ of data series of each model and we have assumed that the correlation matrix ${\bf C_{T_d}}$ of this single realization of each model represents the true correlation matrix of the corresponding system. This approach removes the degeneracy of the $\gamma$-parameters of models and at the same time allows to treat models in a way more similar to the one used for real data. In order to test the stability of filtering procedures with respect to statistical uncertainty (as discussed in subsection \ref{samplesfilt}), we have constructed bootstrap replicas of the single realization ${\bf X_{T_d}}$ of each model. The bootstrap approach has the advantage that it does not require to make assumptions about the data distribution \footnote{We have also used the Cholesky decomposition of ${\bf C_{T_d}}$ instead of the bootstrap approach, in order to obtain different realizations of the non degenerate systems. The Cholesky decomposition approach \cite{johnson94} allows to construct mutually independent realizations of the system. However results obtained with the Cholesky decomposition are in complete agreement with results obtained by using the bootstrap technique that we report in the paper. It is also to notice that by using the Cholesky decomposition to perform simulations it is necessary to know the data distribution (e.g. Gaussian or Student-t), whereas the bootstrap approach does not require to make assumptions about such distribution.}.  

We have simulated $1000$ independent sets of data for the artificial systems described by the degenerate models and we have constructed $1000$ bootstrap replicas \cite{Efron,Tumminello2007} of the empirical data. We have also considered 1000 bootstrap replicas of the single realization with series length $T_d$ of both the artificial systems, in order to treat the models more similarly to real data. We have applied all the filtering procedures described above to the correlation matrix ${\bf C_i}$ of  each simulation or replica $i$ of the artificial systems and to each replica $i$ of the real systems. Therefore, we have obtained four filtered correlation matrices that we indicate with ${\bf C^{filt}_i}$ associated with each realization or replica $i$ of the systems. The label ${\bf filt}$ in ${\bf C^{filt}_i}$ stands for ALCA, SLCA, B and S depending on the filtering procedure.

\subsection{Information about the model}\label{filtmod}

The first question we want to ask is which filtering procedure performs better in detecting the correlation matrix of the model. We can ask this question only for the simulations where we know the model correlation matrix used to generate the data.
In order to evaluate the ability of filtering procedures in reconstructing the correlation matrix of the model ${\bf \Sigma}$, we have evaluated the average Kullback-Leibler distance $\langle K({\bf \Sigma},{\bf C^{filt}_i}) \rangle$ between the correlation matrix of the model and the correlation matrix filtered from the samples. Averages have been performed over $1000$ realizations of the models. The smaller $\langle K({\bf \Sigma},{\bf C^{filt}_i}) \rangle$ the larger is the amount of information about the model that is detected by the filtered matrix. In Tables I and II we distinguish between degenerate models that we indicate with ``Block diagonal'' and ``HNFM'' and non degenerate models that we indicate with  ``Block diagonal (n.d.)'' and ``HNFM (n.d.)''. 
In Table I we report results obtained for all the considered models when the length of simulated normally distributed time series is $T=748$. In the table, we observe that the ALCA outperforms all the other filtering procedures both for degenerate and non degenerate models. It is also to notice that the performance of SLCA is better than both the spectral filtering procedures for all the models with the exception of the non degenerate block diagonal model. Such a good performance of hierarchical clustering filtering procedures was expected for the degenerate models. Indeed, as we have discussed above, such models give a certain advantage to hierarchical clustering filtering procedures because of the degeneracy of coefficients. The fact that ALCA outperforms all the other filtering procedures also in the case of non degenerate models can be explained by taking into account both the length of data series and the way in which model degeneracy has been removed. The correlation matrix of the non degenerate models is by construction the correlation matrix of a single realization of the corresponding degenerate models with series length $T_d=748$. This fact implies that the dispersion of the non degenerate correlations from the corresponding values in the degenerate model is of the order of $D_m=1/\sqrt{T_d}=1/\sqrt{748}$. In Table I, the length of simulated data series is also $T=748$, i.e. $T=T_d$. This fact implies that the statistical uncertainty associated with the sample correlations is of the order $1/\sqrt{T}=1/\sqrt{748}$. This value is equal to $D_m$, implying that for series length $T=748$ the non degeneracy of model parameters is of the same order of the statistical uncertainty. In other words, details about specific correlation values cannot be distinguished from statistical uncertainty for such short data series. Only the global structure of the correlation matrix is important and hierarchical clustering procedures results to be more capable than spectral techniques in reconstructing the correlation structure of the models. In order to better understand the effect of the non degeneracy of model parameters on the ability of filtering procedures in reconstructing the model, we consider also a case with time series of length longer than in the prevoius case. Specifically, in Table II we report results obtained for time series of length $T=7480$, which is ten times the length considered in Table I. In the case of $T=7480$, we continue to observe a better performance of hierarchical clustering filtering procedures and in particular of ALCA with respect to spectral techniques for the degenerate models. This fact was expected because of the degeneracy of the models. However, in Table II we observe that the spectral technique producing ${\bf C^B}$ as result of the filtering outperforms hierarchical clustering procedures for the non degenerate models. 
The method producing ${\bf C^S}$ provides a result which is of the same order than ${\bf C^{ALCA}}$ for the block diagonal (n.d.) model whereas still underperform with respect to both hierarchical clustering procedures for the HNFM (n.d.). The success of ${\bf C^B}$ can be explained by the fact that for $T=7480$ the statistical uncertainty of sample correlations is of the order $1/\sqrt{T}=1/\sqrt{7480}$ which is smaller than $D_m$. Therefore, for $T=7480$ the non degeneracy of models becomes relevant as compared with the statistical uncertainty affecting sample correlations and spectral techniques result to be more capable than hierarchical clustering in taking into account such non degeneracy. This aspect is related to the fact that ALCA and SLCA are filtering procedures characterized by $n-1$ free parameters whereas spectral methods have a variable number of free parameters which is scaling as $n^2$ when T tends to infinity.

In summary, we have shown that hierarchical clustering procedures better reconstruct the degenerate models both for short and long time series, whereas for the non degenerate models the length of data series becomes relevant in the comparison. Specifically, for short time series ($T=748$), such that the statistical uncertainty of correlations hides the heterogeneity of model parameters, we have observed that hierarchical clustering procedures, and in particular the ALCA, outperform spectral techniques. On the contrary, for data series long enough ($T=7480$) that the heterogeneity of model parameters is relevant with respect to the statistical uncertainty of sample correlations, spectral procedures result typically to be more efficient than hierarchical clustering procedures in reconstructing the correlation matrix of models.

\begin{widetext} 
\begin{center}
\begin{table}\nonumber
\begin{minipage}{0.65 \linewidth}
\caption{Average value of the Kullback-Leibler distance between the correlation matrix 
of the model and the correlation matrix filtered from the sample one. For each case average 
and standard deviation are obtained from 1000 realizations or bootstrap replicas of the system.
 ($n=100$, $T=748$).}
\end{minipage}
\begin{tabular}{||c|c|c|c|c||}
\tableline
\tableline
{\bf Models} & $\langle K({\bf \Sigma},{\bf C^{ALCA}_i}) \rangle$ 
& $\langle K({\bf \Sigma},{\bf C^{SLCA}_i}) \rangle$ 
& $\langle K({\bf \Sigma},{\bf C^{B}_i}) \rangle$ 
& $\langle K({\bf \Sigma},{\bf C^{S}_i}) \rangle$ \\
\tableline
\tableline
Block diagonal & $0.15 \pm 0.01$ & $0.57 \pm 0.04$ & $0.84 \pm 0.03$ & $1.50 \pm 0.05$ \\
\tableline
HNFM & $0.22 \pm 0.02$ & $0.33 \pm 0.05 $& $1.99 \pm 0.07$ & $2.15 \pm 0.08$\\
\tableline
Block diagonal (n.d.) & $3.56 \pm 0.02$ & $4.36 \pm 0.07 $ & $3.74 \pm 0.06$ & $4.34 \pm 0.09$\\
\tableline
HNFM (n.d.)& $3.38 \pm 0.02$ & $3.85 \pm 0.08$ & $4.54 \pm 0.08$ & $5.0 \pm 0.1$\\
\tableline
\tableline
\end{tabular}
\end{table}



\begin{table}\nonumber
\caption{The same as in Table I but with $T=7480$.}
\begin{tabular}{||c|c|c|c|c||}
\tableline
\tableline
{\bf Models} & $\langle K({\bf \Sigma},{\bf C^{ALCA}_i}) \rangle$ 
& $\langle K({\bf \Sigma},{\bf C^{SLCA}_i}) \rangle$ 
& $\langle K({\bf \Sigma},{\bf C^{B}_i}) \rangle$ 
& $\langle K({\bf \Sigma},{\bf C^{S}_i}) \rangle$ \\
\tableline
\tableline
Block diagonal & $0.015 \pm 0.001$ & $0.105 \pm 0.006$ & $0.162 \pm 0.006$ & $0.70 \pm 0.01$ \\
\tableline
HNFM & $0.023 \pm 0.002$ & $0.032 \pm 0.005 $& $ 0.986\pm 0.007$ & $1.44 \pm 0.07$\\
\tableline
Block diagonal (n.d.) & $3.418 \pm 0.004$ & $3.94 \pm 0.02 $ & $ 2.95\pm 0.02$ & $3.41 \pm 0.02$\\
\tableline
HNFM (n.d.)& $3.174 \pm 0.008$ & $3.52 \pm 0.02$ & $2.54 \pm 0.04$ & $4.66 \pm 0.09$\\
\tableline
\tableline
\end{tabular}
\end{table}
\end{center}
\end{widetext}

\subsection{Information about the sample correlation matrix and stability}\label{samplesfilt}

In this subsection we quantify the amount of information that different filtering procedures preserve when applied to sample correlation matrices. This is important in all those real cases when one does not know the model correlation matrix. Moreover we investigate the stability of the filtered correlation matrices with respect to different realizations of the process. 
We use two quantities in order to evaluate the performance of the filtering procedures. The first quantity that we have measured is the Kullback-Leibler distance $K({\bf C_i},{\bf C^{filt}_i})$ between the correlation matrix ${\bf C_i}$ of the $i$-th sample  and the filtered correlation matrix ${\bf C^{filt}_i}$ obtained by applying one of the filtering procedure to ${\bf C_i}$. $K({\bf C_i},{\bf C^{filt}_i})$ is a measure of the information about ${\bf C_i}$ that is stored in ${\bf C^{filt}_i}$: the smaller $K({\bf C_i},{\bf C^{filt}_i})$, the larger is the amount of information about ${\bf C_i}$ which is retained in the filtered matrix. The second quantity that we have considered is the Kullback-Leibler distance  $K({\bf C^{filt}_i},{\bf C^{filt}_j})$ between two filtered matrices ${\bf C^{filt}_i}$ and ${\bf C^{filt}_j}$ obtained by applying the same filtering procedure to two different simulations (or replicas) $i$ and $j$ of the system. $K({\bf C^{filt}_i},{\bf C^{filt}_j})$ measures the statistical robustness of filtered matrices. The smaller $K({\bf C^{filt}_i},{\bf C^{filt}_j})$, the greater is the stability of the filtering procedure with respect to the statistical uncertainty. In our estimations, we have averaged both $K({\bf C_i},{\bf C^{filt}_i})$ and $K({\bf C^{filt}_i},{\bf C^{filt}_j})$ over the $1000$ independent realizations or replicas of each system.


\begin{figure}
\begin{center}
\includegraphics[width=0.48\textwidth]{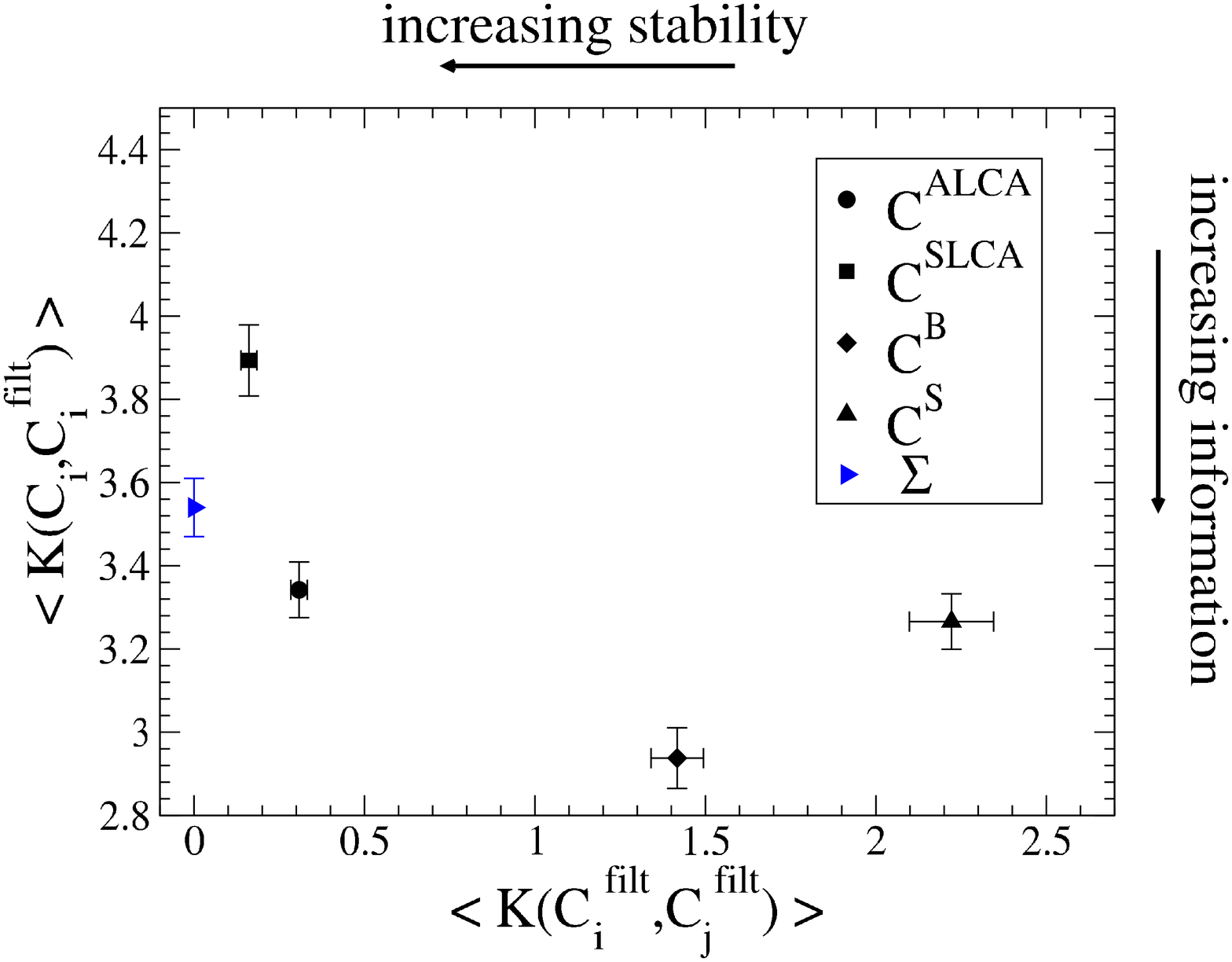}
\caption{Block diagonal model with degenerate coefficients. Comparison of the 4 filtered correlation matrices described in the text. In the graph we plot the stability of the filtered matrix ($x$ axis) against the amount of information about the correlation matrix that is retained in the filtered matrix ($y$ axis). Small values of $\langle K({\bf C^{filt}_i},{\bf C^{filt}_j}) \rangle$ and $\langle K({\bf C_i},{\bf C^{filt}_i}) \rangle$ correspond to large stability and large amount of information preserved by the filtering respectively. The analysis is performed for a system of 100 elements divided in 12 orthogonal groups, each one depending on a specific Gaussian factor, i.e. a block diagonal model. Averages have been performed over 1000 independent realizations of the system and error bars correspond to one standard deviation.}
\label{infostabNOHdeg}
\end{center}
\end{figure}

\begin{figure}
\begin{center}
\includegraphics[width=0.48\textwidth]{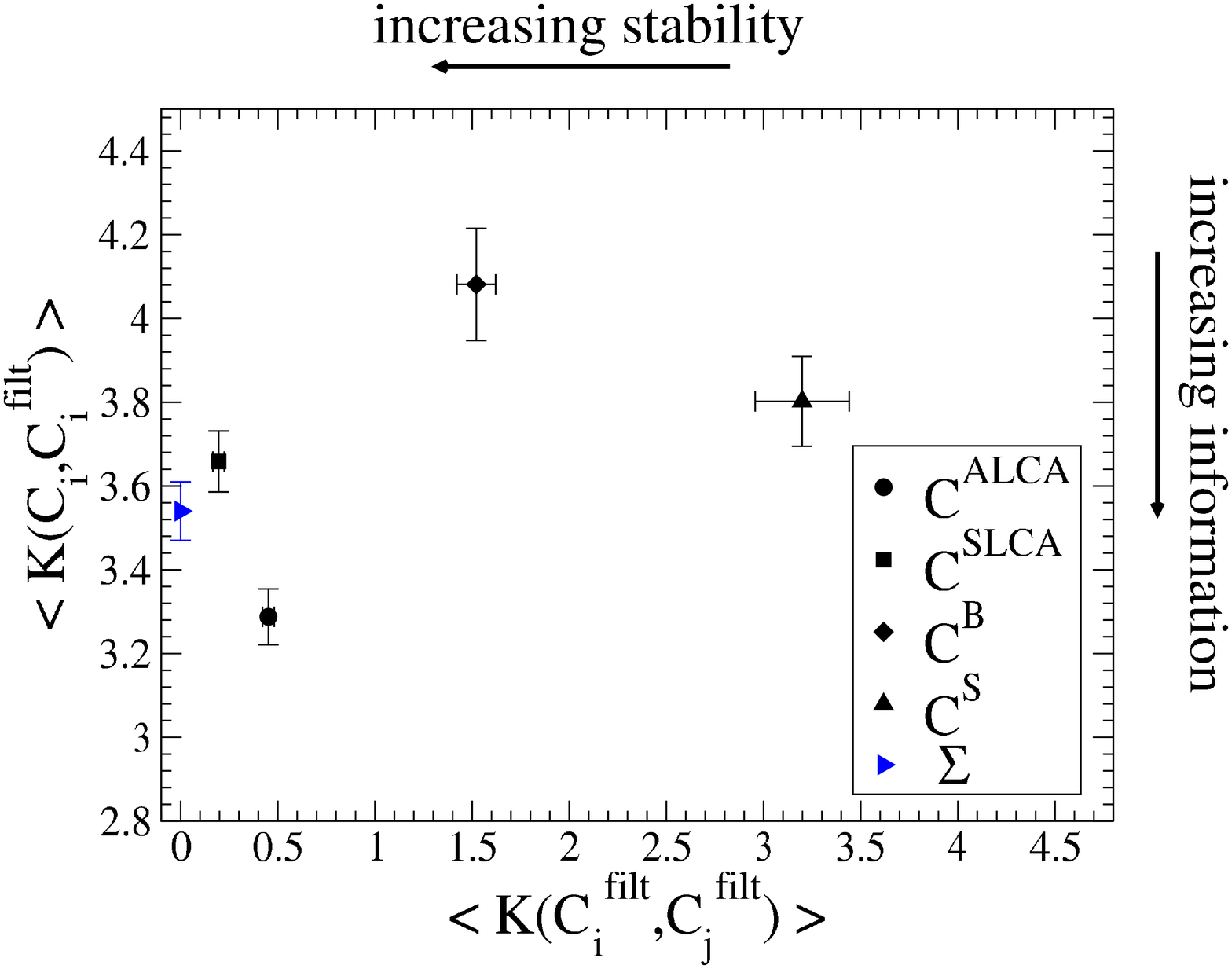}
\caption{Hierarchically nested factor model with degenerate coefficients. Comparison of the filtered correlation matrices produced by the 4 techniques described in the text. The analyzed system is composed by 100 elements following the HNFM with 23 factors obtained in Ref. \cite{TumminelloEPL2007}. Averages have been performed over 1000 independent realizations of the system and error bars correspond to one standard deviation.}
\label{infostabHNFMdeg}
\end{center}
\end{figure}

In Fig. \ref{infostabNOHdeg}, we show the results obtained for the block diagonal model with degenerate coefficients. In the figure we plot $\langle K({\bf C_i},{\bf C^{filt}_i}) \rangle$ versus $\langle K({\bf C^{filt}_i},{\bf C^{filt}_j}) \rangle$ for all the described filtering procedures.  Averages that we indicate with the notation $\langle . \rangle$ are performed over 1000 realizations and the series length is $T=748$. Error bars are one standard deviation.  In all the cases presented in this paper we have verified that the error interval indicated around the mean value of plus and minus one standard deviation includes approximately the 67\% of the realizations used to compute the mean value. In the figure we also report the result of an hypothetic perfect filtering procedure, i.e. a filtering techniques which is able to recover exactly the model from each realization. In the figure, we indicate the corresponding correlation matrix with ${\bf \Sigma}$. Such a filtering is maximally stable, because it recovers always the correlation matrix of the block diagonal factor model. Accordingly, it is $\langle K({\bf \Sigma},{\bf \Sigma}) \rangle=0$. This perfect filtering procedure removes completely the noise due to the finite length of data series and therefore the quantity $\langle K({\bf C_i},{\bf \Sigma}) \rangle \neq 0$. Instead, it is equal to the expectation value of Eq. (\ref{kullexpecS1Sig}), i.e. $\langle K({\bf C_i},{\bf \Sigma}) \rangle \simeq 3.54$ for $n=100$ and $T=748$. Note that we know the position in the plane of the optimal filtering even if we do not know the underlying model. This is due to the important characteristic that the mean value of the Kullback-Leibler distance is independent from the model correlation matrix (at least in the multivariate Gaussian case).
In the figure, we observe that all the filtering procedures, except the SLCA, retain in average more information about the sample correlation matrix than the true model, i.e. $\langle K({\bf C_i},{\bf C^{filt}_i}) \rangle < 3.54$ for ${\bf C^{filt}}$ equal to ${\bf C^{ALCA}}$, ${\bf C^{B}}$ and ${\bf C^{S}}$. This fact indicates that these filtering procedures do not discard completely the noise present in the sample correlation matrix as a consequence of the finite length of time series. The SLCA algorithm is the only one which is retaining less information about the sample correlation matrix than the true model. Moreover the SLCA is more stable than all the other filtering procedures.

In Fig. \ref{infostabHNFMdeg}, we show the results obtained by applying the considered filtering procedures to the system described by the HNFM with $P=23$ factors and with degenerate coefficients. In this case, only the ALCA is retaining more information about the sample correlation matrix than the true model. However it is interesting to note that both the spectral techniques are at the same time less informative about the sample correlation matrix and less stable than both hierarchical clustering filtering procedures. In other words, for the degenerate HNFM, hierarchical clustering procedures clearly outperform spectral techniques. This fact is a consequence of the pure hierarchical nature of the HNFM. Indeed in Ref. \cite{TumminelloEPL2007}, we have shown that when the hierarchical features of a system are prominent with respect to the details of specific correlation values, the spectral procedures have problems in filtering information about the system. Such problems do not appear for separable systems, like the block diagonal model considered above.

In summary, for both the considered models we observe that hierarchical clustering techniques produce more stable filtered correlation matrices than spectral procedures. Concerning the information about the sample correlation matrix that is stored in the filtering we observe
that results obtained for hierarchical clustering procedures are closer to the perfect filtering (giving as output the true model of the system) than spectral techniques. Finally, it is to notice that the SLCA is the most stable within the considered filtering procedures. Such an excellent performance of hierarchical clustering techniques can be due to the degenerate nature of models as discussed in the first part of this section.

\begin{figure}
\begin{center}
\includegraphics[width=0.48\textwidth]{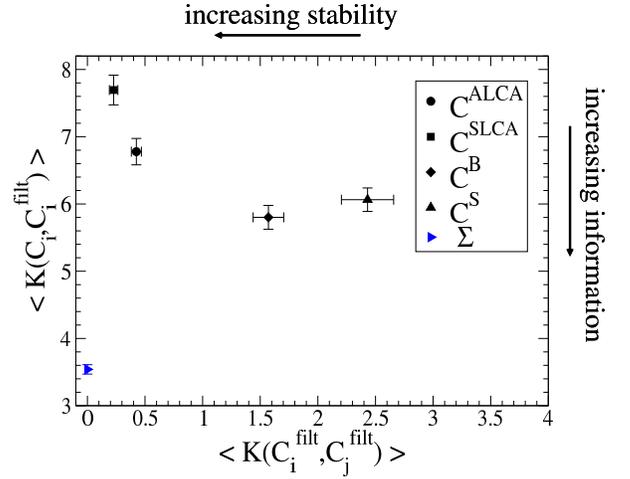}
\caption{Block diagonal model with non degenerate coefficients. Comparison of the 4 filtered correlation matrices described in the text. 
The analysis is performed for a system of 100 elements divided in 12 orthogonal groups, each one depending on a specific Gaussian factor, i.e. a block diagonal model. Averages have been performed over 1000 bootstrap replicas of a single realization of the system and error bars correspond to one standard deviation.}
\label{infostabNOH}
\end{center}
\end{figure}

\begin{figure}
\begin{center}
\includegraphics[width=0.48\textwidth]{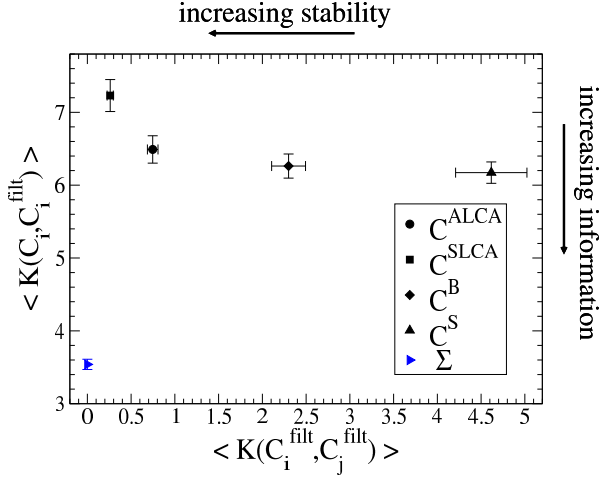}
\caption{Hierarchically nested factor model with non degenerate coefficients. Comparison of the filtered correlation matrices produced by the 4 techniques described in the text. 
The analyzed system is composed by 100 elements following the HNFM with 23 factors obtained in Ref. \cite{TumminelloEPL2007}. Averages have been performed over 1000 bootstrap replicas of a single  realization of the system and error bars correspond to one standard deviation.}
\label{infostabHNFM}
\end{center}
\end{figure}

In fact when we remove the degeneracy of coefficients from the models we observe a different behavior of filtering procedures.
In Fig. \ref{infostabNOH} we plot $\langle K({\bf C_i},{\bf C^{filt}_i}) \rangle$ versus $\langle K({\bf C^{filt}_i},{\bf C^{filt}_j})\rangle$  for the artificial system obtained from a single realization ${\bf X_{T_d}}$ with time series length $T_d=748$ of the factor model with 12 orthogonal factors. This is equivalent to consider a block model with non degenerate coefficients.
In Fig. \ref{infostabHNFM}, we plot results obtained for the single realization with length $T_d=748$ of time series of the HNFM with 23 factors. Also in this case our investigation is equivalent to consider a HNFM with non degenerate coefficients. 
%
%
Mean values and error bars in the figures correspond to the average and the standard deviation respectively both estimated over 1000 bootstrap replicas of the single realization of the models. From Figures \ref{infostabNOH} and \ref{infostabHNFM} we note that 
\begin{eqnarray}\label{infofilered}
\langle K({\bf C_i},{\bf C^{B}_i})\rangle\, \simeq \, \langle K({\bf C_i},{\bf C^{S}_i})\rangle \, \lesssim \,  \nonumber \\ 
\, \lesssim \, \langle K({\bf C_i},{\bf C^{ALCA}_i})\rangle\, \lesssim \,\langle K({\bf C_i},{\bf C^{SLCA}_i})\rangle \nonumber
\end{eqnarray} 
In both the figures, we observe that none of the filtering procedures is more informative about the sample correlation matrix than the true correlation matrix ${\bf \Sigma}={\bf C_{T_d}}$ of both the models, i.e. $E \left[K({\bf C},{\bf \Sigma}) \right]\simeq 3.54$ is smaller than any $\langle K({\bf C_i},{\bf C^{filt}_i})\rangle$ reported in the figures. 

Concerning the stability of the filtered matrices, from the figures we observe that the SLCA filtered matrix outperforms all the other techniques, although the filtered matrix given by ALCA has a stability of the same order of magnitude of the SLCA matrix. 
A good filtered correlation matrix should be at least more stable than the sample correlation matrix with respect to the statistical uncertainty. This sentence can be translated in the following inequality
\begin{equation}\label{stabfilt}
\langle K({\bf C^{filt}_i},{\bf C^{filt}_j}) \rangle < \langle K({\bf C_i},{\bf C_j}) \rangle.
\end{equation}  
For Gaussian variables we know the expected value of $K({\bf C_i},{\bf C_j})$ from Eq. (\ref{kullexpecS1S2}) and thus, for $n=100$ and $T=748$, the last inequality becomes
\begin{equation}\label{compstab}
\langle K({\bf C^{filt}_i},{\bf C^{filt}_j}) \rangle < \frac{1}{2} \frac{n (n+1)}{T-n-1}\, \simeq 7.81.
\end{equation}
This condition is satisfied by all the considered filtered matrices. However we stress the fact that the matrices obtained from hierarchical clustering techniques and in particular the one obtained by SLCA have a value of $\langle K({\bf C^{filt}_i},{\bf C^{filt}_j}) \rangle$ of an order of magnitude smaller than the one expected for the Pearson estimator of correlations. 

In summary, our investigation of considered models shows that spectral filtering techniques are slightly more informative about the sample correlation matrix than hierarchical clustering filtering techniques when details about specific correlation values are relevant, like in the case of non degenerate models. On the contrary, from the point of view of stability of filtered matrices, hierarchical clustering procedures, and in particular the SLCA, outperform spectral techniques.

\subsection{Empirical data}\label{empirical}

In this subsection, we compare the filtering procedures when applied to real data. We have considered the system of daily returns of the 100 most capitalized stocks traded at NYSE in the time period from January 2001 to December 2003. In this case, the length of the $n=100$ time series is $T=748$ records. We have also considered the system of daily returns of 92 highly capitalized stocks traded at London Stock Exchange in 2002. For this system the record length of the $n=92$ time series is $T=250$.

\begin{figure}
\begin{center}
\includegraphics[width=0.48\textwidth]{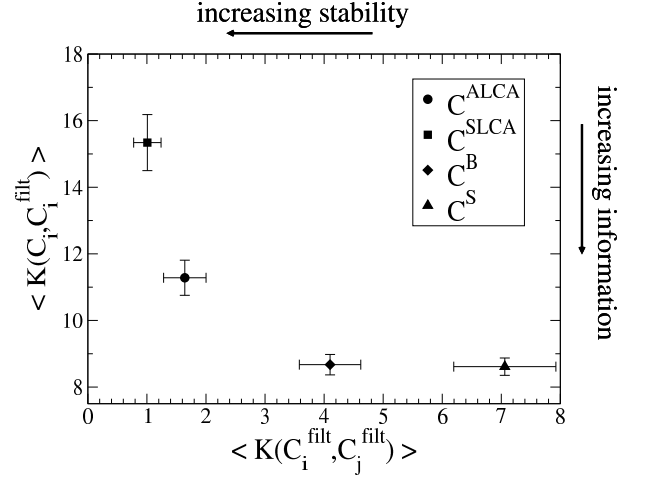}
\caption{Correlation matrix of a real system composed of 100 stocks traded at NYSE during the period from January 2001 to December 2003. The variable investigated is daily return of the most capitalized stocks. The length of time series is $T=748$ for this system. Averages have been performed over 1000 bootstrap replicas of data series and error bars correspond to one standard deviation.}
\label{infostabNYSE100}
\end{center}
\end{figure}

\begin{figure}
\begin{center}
\includegraphics[width=0.48\textwidth]{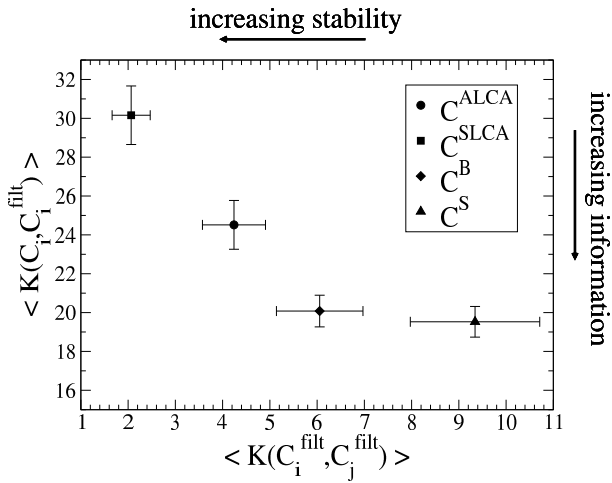}
\caption{Correlation matrix of a real system composed of 92 stocks traded at LSE during the period from January to December 2002. The variable investigated is daily return of the most capitalized stocks. The length of time series is $T=250$ for this system.  Averages have been performed over 1000 bootstrap replicas of data series and error bars correspond to one standard deviation.}
\label{infostabLSE92}
\end{center}
\end{figure}

In Fig. \ref{infostabNYSE100}, we report the results obtained by applying all the considered filtering procedures to the system of $n=100$ stocks traded at NYSE, while in Fig. \ref{infostabLSE92} we show the results obtained for the system of $n=92$ stocks traded at LSE. In both the figures, we observe that hierarchical clustering procedures are more stable than spectral techniques, whereas the latter are more informative about the sample correlation matrix than hierarchical clustering. These facts are in agreement with results obtained for simulations in the case of non degenerate models. However this agreement is only qualitative. Indeed, both the values of  $\langle K({\bf C_i},{\bf C^{filt}_i}) \rangle$ and $\langle K({\bf C^{filt}_i},{\bf C^{filt}_j}) \rangle$ observed for the real systems are larger than the corresponding values obtained in the case of simulations. This fact can be due to two effects. The first one is related to the fact that the real systems can be characterized by a structure of correlations more complex than the one considered in the models. For example, the role of the complexity of correlation structures onto the performance of filtering procedures was observed in the simulations of the degenerate models of subsection \ref{filtmod} for the spectral techniques. Indeed the performance of such procedures was rather unsatisfactory for the HNFM with respect to the block diagonal model. The second effect that can be responsible for the quantitative difference between results obtained for simulations and results obtained for real data can be related to the fact that we have considered Gaussian variables in the simulations, whereas the distribution of returns is fat tailed \cite{MS00}.

Some quantitative differences are also evident in the comparison of the two real systems. Specifically, both the values of  $\langle K({\bf C_i},{\bf C^{filt}_i}) \rangle$ and $\langle K({\bf C^{filt}_i},{\bf C^{filt}_j}) \rangle$ are larger in the LSE data with respect to the NYSE data. This difference is mainly due to the different length of data series, i.e. $T=748$ at NYSE and $T=250$ at LSE. The smaller $T$, the larger is the statistical uncertainty of the sample correlation matrix. For instance, we can make quantitative this difference by using the expectation values of the Kullback-Leibler distance of Eq.s (\ref{kullexpecS1Sig}) and (\ref{kullexpecS1S2}). For a system of $n=100$ elements with data series of length $T=748$ we have $E\left[K({\bf C_1},{\bf \Sigma})\right]\simeq3.54$ and $E\left[K({\bf C_1},{\bf C_2})\right]\simeq7.81$, whereas for a system of $n=92$ elements and series length $T=250$ is $E\left[K({\bf C_1},{\bf \Sigma})\right]\simeq9.86$ and $E\left[K({\bf C_1},{\bf C_2})\right]\simeq27.2$. 
A comparison of the results obtained for Gaussian random models in subsection \ref{samplesfilt} with the results obtained for the real systems investigated in this subsection shows that the Kullback-Leibler distance provides results on real data about the relative effectiveness of the considered filtering procedures which are in agreement with those observed for models.

\section{Conclusions}\label{conclusions}

In conclusion we have shown that the Kullback-Leibler distance can be fruitfully used to compare correlation matrices of multivariate data. We have shown that this distance is more appropriate to achieve this objective than the standard Frobenius distance. This fact is due to some properties of the Kullback-Leibler distance such as the asymmetry, the model independence of expectation values and its relation with the maximum likelihood factor analysis. 
Sample correlation matrices can be compared in pairs among them and/or with respect to model matrices or to filtered matrices.   
We have used the Kullback-Leibler distance to compare four different techniques used to obtain a filtered correlation matrix from the empirical one. Two of the four techniques that we have analyzed are spectral filtering procedures based on random matrix theory whereas the other two techniques are based on hierarchical clustering methods, specifically ALCA and SLCA. Results obtained for simulations are consistent with those obtained for real data. These results can be summarized as follows: both the considered spectral techniques are slightly more informative about the sample correlation matrix than the other two techniques based on hierarchical clustering. On the other hand both the techniques based on hierarchical clustering are producing filtered correlation matrices which are more stable than those obtained with spectral procedures. 
These results show that the Kullback-Leibler distance is very useful in characterizing multivariate systems described by real data, factor models and matrices filtered from the sample one. 
%
%

In conclusion, the Kullback-Leibler distance is a powerful and accurate tool able to characterize the information and stability of sample, model and filtered correlation matrices and it is a useful quantitative indicator for the relative amount of information and the relative stability of correlation matrices of multivariate data.  

\section{Acknowledgements}

We acknowledge partial support from MIUR research project ``Dinamica di altissima frequenza nei mercati finanziari'' and NEST-DYSONET 12911 EU project.

\section{Appendix A}
In this Appendix, we show how to derive Eq. (\ref{kullbackGaussian}) from Eq. (\ref{kullbackMulti}). Let us consider the multivariate Gaussian distributions $P({\bf \Sigma_1},X)$ and $P({\bf \Sigma_2},X)$ describing the same random vector $X$. We have
\begin{equation}
\label{multigauss}
P({\bf \Sigma_i},X)=\frac{1}{\sqrt{(2 \pi)^n \abs{{\bf \Sigma_i}}}} \exp\left (- \frac{1}{2} X^T {\bf \Sigma_i}^{-1} X \right ).
\end{equation}
By substituting Eq. (\ref{multigauss}) into Eq. (\ref{kullbackMulti}) we get
\begin{eqnarray}
\label{pass1kull}
K(P({\bf \Sigma_1},X),P({\bf \Sigma_2},X))=\frac{1}{2}\log \left( \frac{\abs{{\bf \Sigma_2}}}{\abs{{\bf \Sigma_1}}}\right )+ \nonumber\\
+\frac{1}{2 \sqrt{(2 \pi)^n \abs{{\bf \Sigma_1}}} }(I_{1,2}-I_{1,1}),
\end{eqnarray}
where 
\begin{equation}
\label{Iij}
I_{i,j}=\int{e^{-\frac{1}{2} X^T {\bf \Sigma_i}^{-1} X} \left (X^T {\bf \Sigma_j}^{-1} X \right )dX}.
\end{equation}

The integral $I_{i,j}$ can be solved by using the linear transformation $Y={\bf G_j} X$, where ${\bf G_j}$ is the orthogonal matrix which diagonalizes ${\bf \Sigma_j}$. It results that
\begin{equation}
\label{pass2kull}
I_{i,j}=\sqrt{(2 \pi)^n \abs{{\bf \Sigma_i}}} \sum_{q=1}^{n} h_{qq} b_{qq},
\end{equation}
where $h_{qq}$ $(q=1,...,n)$ are the elements of the diagonal matrix ${\bf G_j}^T {\bf \Sigma_j}^{-1} {\bf G_j}$, whereas $b_{qq}$ $(q=1,...,n)$ are the diagonal elements of the matrix ${\bf G_j}^T {\bf \Sigma_i} {\bf G_j}$. We can further simplify the expression of $I_{i,j}$ by taking into account the fact that the matrix ${\bf G_j}^T {\bf \Sigma_j}^{-1} {\bf G_j}$ is diagonal. Indeed $\sum_{q=1}^{n} h_{qq} b_{qq}=\text{tr}[{\bf G_j}^T {\bf \Sigma_j}^{-1} {\bf G_j} {\bf G_j}^T {\bf \Sigma_i}{\bf G_j}]$ $=$ $\text{tr}[{\bf \Sigma_j}^{-1}{\bf \Sigma_i}]$ due to the orthogonality of ${\bf G_j}$ and to the invariance of the trace with respect to rotations. Accordingly, we obtain that
\begin{equation}
\label{pass3kull}
I_{i,j}=\sqrt{(2 \pi)^n \abs{{\bf \Sigma_i}}} \,\,\,\, \text{tr}[{\bf \Sigma_j}^{-1}{\bf \Sigma_i}].
\end{equation}

Finally, we obtain the expression of $K(P({\bf \Sigma_1},X),P({\bf \Sigma_2},X))$ given in Eq. (\ref{kullbackGaussian}) by substituting the last expression of $I_{i,j}$ into Eq. (\ref{pass1kull}) and noting that $ \text{tr}[{\bf \Sigma_i}^{-1}{\bf \Sigma_i}]=n$.\\

\section{Appendix B}

In this Appendix, we derive the expectation values of the Kullback-Leibler distance given in Eq.s (\ref{kullexpecSigS1}-\ref{kullexpecS1S2}). We shall use two known results from the theory of Wishart matrices. Let us consider a multinormally distributed random vector $X$ of dimension $n$ with correlation matrix ${\bf \Sigma}$. Let ${\bf C_1}$ and ${\bf C_2}$ be two sample correlation matrices obtained from two independent realizations of the system, ${\bf X}_1$ and ${\bf X}_2$ respectively both of length $T$. The first result from the theory of Wishart matrices that we shall use hereafter is that $\log \abs{{\bf C_i}}$, $i=1,2$ is equal to $\log \abs{{\bf \Sigma}}-n \log(T)$ plus the sum of the logarithms of $n$ mutually independent chi-squared random variables $y_{T-n+1},...,y_{T}$ with degrees of freedom $T-n+1,...,T-1,T$ respectively (see for instance \cite{Mardia}). This fact implies that the expectation value of $\log \abs{{\bf C_i}}$ is
\begin{equation}
\label{logexpec}
E(\log \abs{{\bf C_i}})=\log \abs{{\bf \Sigma}}-n \log(T)+\sum_{p=T-n+1}^T E[\log(y_p)].
\end{equation}
Because $E[\log(y_p)]=\Gamma^{\prime}(p/2)/\Gamma(p/2)+\log(2)$ (see for instance \cite{integrals}) we obtain that:
\begin{equation}
\label{logexpecSEC}
E(\log \abs{{\bf C_i}})=\log \abs{{\bf \Sigma}}+n \log(2/T)+\sum_{p=T-n+1}^T \frac{\Gamma^{\prime}(p/2)}{\Gamma(p/2)}.
\end{equation}

The other result from the theory of Wishart matrices that we use here is that the expectation value of the inverse of ${\bf C_i}$ is $E({\bf C_i}^{-1})=T {\bf \Sigma}^{-1}/(T-n-1)$ (see for instance \cite{Mardia}). Accordingly we obtain: 

\begin{eqnarray}
\label{expectrace}
E\left [\text{tr}\left({\bf C_i}^{-1} {\bf \Sigma}\right)\right]=E\left [\text{tr}\left({\bf C_i}^{-1} {\bf C_j}\right)\right]=  \frac{n \, T}{T-n-1}, 
\end{eqnarray}

where we have used the linearity of the trace operator. Finally, we have:

\begin{eqnarray}
\label{expectraceSigmaS1}
E\left[ \text{tr} \left({\bf \Sigma}^{-1} {\bf C_j}\right)\right]=\text{tr}\left({\bf \Sigma}^{-1} {\bf \Sigma} \right)=n, 
\end{eqnarray}

where we have again used the linearity of the trace and the fact that $E({\bf C_i})={\bf \Sigma}$. By using Eq.s (\ref{logexpecSEC}) and (\ref{expectrace}) it is now straightforward to obtain both the expression of $E\left[K({\bf \Sigma},{\bf C_1})\right]$ as given in Eq. (\ref{kullexpecSigS1}) and the expectation value $E\left[K({\bf C_1},{\bf C_2})\right]$ as given in Eq. (\ref{kullexpecS1S2}). Finally, by using results of Eq.s (\ref{logexpecSEC}) and (\ref{expectraceSigmaS1}) we obtain the expectation value of $K({\bf C_1},{\bf \Sigma})$ as given in Eq. (\ref{kullexpecS1Sig}).


\end{document}